\documentclass[12pt]{amsart}
\usepackage{fullpage}
\usepackage{graphicx} 
\usepackage{amsfonts}
\usepackage{amsthm}
\usepackage{amsmath}
\usepackage{amssymb}
\usepackage{mathabx}
\usepackage{mathrsfs}
\usepackage{mathptmx}
\newtheorem{theorem}{Theorem}
\newtheorem{lemma}{Lemma}

\newtheorem{corollary}{Corollary}
\theoremstyle{definition}


\begin{document}

\title{On the accuracy of population level approximation of network processes}
\date{}
\author{No\'emi Nagy}
\address[N. Nagy]{Department of Analysis and Operations Research, Institute of Mathematics,Budapest University of Technology and Economics, Muegyetem rkp. 3., H-1111 Budapest, Hungary}
\address[N. Nagy, S. Horváth]{National Laboratory for Health Security, Budapest, Hungary}
\address[B. Maga, P. L. Simon]{HUN-REN Alfr\'ed R\'enyi Institute of Mathematics, Re\'altanoda street 13-15, H-1053 Budapest, Hungary}
\address[S. Horváth, P. L. Simon]{Institute of Mathematics, ELTE E\"otv\"os Lor\'and University, 
P\'azm\'any P\'eter S\'et\'any 1/c, H-1117 Budapest, Hungary}

\author{S\'andor Horv\'ath}

\author{Bal\'azs Maga}

\author{P\'eter L. Simon}

\maketitle

\begin{abstract}
The individual-based model of simple contagion processes is considered on regular graphs. This model explicitly incorporates the adjacency matrix of the network enabling us to study the effect of network structure on the dynamic of the propagation process. While the asymptotic behaviour of the model is well known, the transient behaviour has been less studied. Our goal in this paper is to give a theoretical estimate on the accuracy of the one-dimensional population-level approximation. This is carried out for arbitrary simple contagion processes and regular Tur\'an graphs.  Numerical evidence is shown that the theoretical estimate is rather sharp for dense graphs.
\end{abstract}
\section{Introduction}

Complex systems, consisting of many similarly behaving components connected in a possibly complicated structure, arise in several areas of science. Significant  research has been devoted to understand how processes evolve on networks and describe them by differential equations. Widely studied propagation processes are epidemic spread on networks of humans or animals, activity propagation in neuronal networks or information dissemination and opinion formation on social networks.  Network processes, in our terminology, are dynamic processes occurring at the nodes of a graph, with the status of a node changing in time. At a given time, the status of a node changes according to the effect of its neighbours.
	
One of the most widely studied processes is epidemic propagation on networks, in the case of which we present the topic and result of this paper. Our investigations will be carried out for general simple contagion processes that will be introduced in the next section.  Signal propagation and resilience of general network processes are dealt with in the recent review papers \cite{Artime, JiYe}, where different models from epidemiology through social sciences and technology to neuronal dynamics are studied.
	
There are several differential equations that are used for modelling the spread of infection \cite{KissMillerSimon, Sharkey}. It is common in all models that the contact structure is represented by a graph, the nodes of which are individuals and its edges are potentially infectious contacts. The vertices can be in two or three different states: susceptible, infected/infectious or recovered. Their number at time $t$ is denoted by $S(t)$, $I(t)$ and $R(t)$. A susceptible node can become infected because of its infectious neighbours and an infected node can move to the susceptible state (in this case the epidemic is called $SIS$ type) or it can move to the recovered state ($SIR$ epidemic). Three main types of differential equation models of epidemic spread are population level, degree-based and individual-based models \cite{FuSmallChen, HouseKeeling2011, KissMillerSimon, pastor2001PRE, PorterGleeson}. The simplest population level system is the mean-field model, which takes the following form for $SIS$ propagation on a network
\begin{equation}
\dot{[I]} = \tau [SI] -\gamma [I], \label{MF}
\end{equation}
where $[I](t)$ is the expected number of infected nodes and $[SI](t)$ is the expected number of $SI$ edges at time $t$, $\tau$ is the rate of infection across an edge and $\gamma$ is the rate of recovery of a node. This differential equation is not closed in the sense that further differential or algebraic equations are needed for the number of $SI$ edges. The widely used combinatorial-based approximation $[SI]\approx d[S][I]/N$ expresses that a susceptible node in a graph with average degree $d$ has $d$ neighbours on average and the proportion of infected nodes among them is $I/N$, where $N$ denotes the number of nodes in the graph. An approximation of this kind is referred to as moment closure, see \cite{Kuehn} for a review about closures. This closure approximation leads to the well-known compartmental model, or so-called mean-field model
\begin{equation}
\dot{\overline{I}} = \tau d\Big(1-\frac{\overline{I}}{N}\Big)\overline{I}-\gamma \overline{I}, \label{MF_NIMFA}
\end{equation}
The corresponding individual-based differential equation is
\begin{equation} \label{NIMFA0}
\dot{x}_i=\tau (1-x_i)(Gx)_i-\gamma x_i,  \quad i=1,\ldots N,
\end{equation}
where $x_i(t)$ approximates  the probability that node $i$ is infected at time $t$ and $G=(g_{ij})$ is the adjacency matrix of the graph. We will refer to this differential equation as individual-based or NIMFA model, abbreviating N-intertwined mean-field approximation \cite{Mieghem,van2009virus}. We note that the notion "mean-field" is used in several contexts, hence we prefer here to use "individual-based model" for \eqref{NIMFA0} and one-dimensional or population level approximation for \eqref{MF_NIMFA}. Moreover, we will use the notation
\begin{equation}
I(t)= \sum_{i=1}^{N} x_i(t) . \label{eq:It}
\end{equation}
for the main observed quantity, the expected number of infected nodes (in other words, the prevalence) obtained from the individual-based model.
	
The theory of formulating closed systems of ODEs describing epidemic spread on networks is well developed, see e.g. \cite{FuSmallChen, HouseKeeling2011, KissMillerSimon, pastor2001PRE, PorterGleeson}, where second order approximations are also dealt with. We chose the individual-based model \eqref{NIMFA0} because it enables us to study the effect of network structure directly starting from the adjacency matrix.
	
The study of the dynamical behaviour of the ODE system \eqref{NIMFA0} incorporates the following main directions of investigation:
	
\begin{itemize}
\item Existence and stability of steady states. This has been completely characterized  in the literature \cite{BodoSimon, KissMillerSimon, KuehnMolter, van2009virus}. The results are based on the fact that \eqref{NIMFA0} is a monotone dynamical system \cite{Smith} yielding global stability of steady states and that periodic orbits cannot occur.
		
\item Dependence of the endemic steady state on the graph parameters. This is known only for regular graphs.
		
\item Transient behaviour of the system. The goal is to derive a coarse grained low dimensional system that determines the total infection $I(t)$ with high accuracy.
		
\item Identify those characteristic properties of the network (like average degree, diameter, leading eigenvalue of the adjacency matrix) that play crucial role in determining the dynamics of the propagation.
\end{itemize}
	
In this paper, we focus on the third question, the low-dimensional approximation, not only for SIS epidemic but also for a general class of processes called simple contagion. We derive a theoretical estimate for the difference of the individual-based system and its population-level approximation, which is the first result in this direction according to our knowledge. Comparison of individual and population-level models has been carried out only numerically, see e.g. the review articles \cite{PrasseVanMieghem,WuytsSieber}.
	
The paper is structured as follows. The individual-based model of a general simple contagion on an arbitrary network and its population-level approximation is derived in Section 2. Regular Tur\'an graphs are introduced in Section 3 together with the derivation of an error term $H$ between the individual and population-level differential equations in Section \ref{sec:errorterm}. The estimate for the error term is derived in Section \ref{sec:error_estim}.  Our main result, the theoretical estimate for the difference of the full model and the approximation is derived in Section \ref{sec:theor_estim}. The sharpness of this estimate is illustrated numerically for dense graphs in Section \ref{sec:sharpness}. The results are summarized and further research directions are shown in the Discussion in Section 6.
	
The main novelty of the paper is that a class of graphs, regular Tur\'an graphs, is identified when a theoretical upper bound can be derived for the difference of the individual-level model and its population-level approximation. This is based on the exponential estimate of the error term $H$, which holds for regular Tur\'an graphs but seems to fail even for slight perturbations of these graphs. According to our knowledge this is the first result when the difference of the full system of differential equations of a network process is theoretically compared to its coarse-grained, low-dimensional approximation.

\section{General model for simple contagion on networks}
	
We consider a general class of dynamic processes that can spread on a network. The network is given by a graph $G$ with $N$ nodes, each of which can be in two different statuses $A$ and $B$. (We note that our approach can be directly extended to process with more than two node statuses, however for simplicity we limit our investigation for the binary case with two statuses.) The transition between the two statuses at a node depends on the status of the node itself and the statuses of its neighbours.  We assume that the transition of a node occurs independently of others and that each neighbour acts independently of all others to induce a transition. Thus the total transition rate is simply the sum of three independent rates:
$$
\mbox{rate of transition } A \to B \ : \ r_{AB}^0 +  r_{AB}^1 n_A +  r_{AB}^2 n_B ,
$$
where $n_A$ and $n_B$ denote the number of neighbours in status $A$ and $B$, respectively, and $r_{AB}^i$ are non-negative constants, for more detailed model description see Section 2.3.2 in \cite{KissMillerSimon}. Similarly, for the opposite transition we have:
$$
\mbox{rate of transition } B \to A \ : \ r_{BA}^0 +  r_{BA}^1 n_A +  r_{BA}^2 n_B .
$$
Processes, for 	which all transitions take this linear form are called simple contagions, in contrast to complex contagions, when the rate is not a linear function of the number of neighbours.
	
One of the most widely used models and the main motivation of our study in this paper is  SIS epidemic propagation, when each node is either infected $A=I$ or susceptible $B=S$ and there are two transitions: infection ($S\to I$)  with rate $\tau n_I$ and recovery with rate $\gamma$. That is $r_{AB}^0=\gamma$ and $r_{BA}^1 =\tau$ while all other rates are equal to zero $r_{AB}^1=r_{AB}^2=0=r_{BA}^0=r_{BA}^2$. In other words, the infection rate depends linearly on the number of infected neighbours and the recovery rate is a constant independently of the number of different neighbours. Section 2.3.2 in \cite{KissMillerSimon} presents several other examples like propagation of neuronal activity in neuronal networks or opinion dynamics described by the voter model. These spreading processes can be specified by choosing the rate constants $r_{AB}^i$ and $r_{BA}^i$ appropriately.
	
Simple contagions can be described by a system of ordinary differential equations in terms of the variables $\langle A_k \rangle (t)$ and $\langle B_k \rangle (t)$ denoting the probability that node $k$ is in status $A$ or $B$ at time $t$. We note that in the case of binary processes we need only $\langle A_k \rangle (t)$ since $\langle B_k \rangle  = 1-\langle A_k \rangle $, because a node at a given time is either in status $A$ or $B$. Using the formulas for the transition rates above, the differential equation for $\langle A_k\rangle$ takes the form
\begin{align*}
    \dot{\langle A_k\rangle} = r_{BA}^0 \langle B_k \rangle - r_{AB}^0 \langle A_k \rangle  &+ r_{BA}^1 \sum_{l=1}^N g_{kl}  \langle B_k A_l \rangle + r_{BA}^2 \sum_{l=1}^N g_{kl}  \langle B_k B_l \rangle \\
    &- r_{AB}^1 \sum_{l=1}^N g_{kl}  \langle A_k A_l \rangle - r_{AB}^2 \sum_{l=1}^N g_{kl}  \langle A_k B_l \rangle ,
\end{align*}
where $(g_{kl})$ is the adjacency matrix of the network $G$ and $\langle X_k Y_l \rangle$ is the joint probability that node $k$ is in status $X$ and node $l$ is in status $Y$ with $X,Y \in \{ A, B\}$. For a detailed explanation of the terms in the differential equation see Section 3.2.1 in \cite{KissMillerSimon} for the case of SIS epidemic propagation. Here, we only explain the meaning of two terms for the Reader's convenience, namely $ r_{BA}^0 \langle B_k \rangle$ expresses the spontaneous (neighbour-independent) transition rate of node $k$ from status $B$ to status $A$, while $r_{BA}^1 \sum_{l=1}^N g_{kl}  \langle B_k A_l \rangle$ is the rate that node $k$ moves from status $B$ to $A$ due to the pressure of $A$ neighbours along edges $g_{kl}$. At this point we draw attention to the fact that this transition depends on the joint probability $ \langle B_k A_l \rangle$, hence the system above is not self-contained. In order to get a self-contained system the independence approximation $\langle X_k Y_l \rangle \approx \langle X_k \rangle \langle Y_l \rangle$ is widely applied. see e.g. Section 3.4.1 in \cite{KissMillerSimon}. This approximation leads to the so-called individual-based model of simple contagions in the form
\begin{align}
\dot{a}_k = r_{BA}^0 b_k - r_{AB}^0 a_k + r_{BA}^1 \sum_{l=1}^N g_{kl} b_ka_l + r_{BA}^2 \sum_{l=1}^N g_{kl} b_kb_l \\ - r_{AB}^1 \sum_{l=1}^N g_{kl} a_ka_l - r_{AB}^2 \sum_{l=1}^N g_{kl} a_kb_l ,
\label{eq:ak_general}
\end{align}
where we used the new notations
$$
a_k \approx \langle A_k \rangle, \quad b_k \approx \langle B_k \rangle, \quad  a_k b_l \approx \langle A_kB_l \rangle ,
$$
expressing that the approximation $a_k$ is not identical to the original variable $\langle A_k \rangle$.
Using that $b_l=1-a_l$ the sums containing $b_l$ can be transformed as
$$
\sum_{l=1}^N g_{kl} b_l = d_k -\sum_{l=1}^N g_{kl} a_l ,
$$
where $d_k= \sum_{l=1}^N g_{kl} $ denotes the degree of node $k$. In the rest of the paper, we will focus on regular graphs, hence now we assume that $d_k=d$ for all $k$. Then differential equation \eqref{eq:ak_general} takes the form
\begin{equation}
\dot{a}_k = \alpha^{+} (1-a_k) - \alpha^{-} a_k + \left( \beta^{+} (1-a_k) - \beta^{-} a_k \right) \sum_{l=1}^N g_{kl} a_l,
\label{eq:ak_regular}
\end{equation}
where the new parameters are
\begin{equation}
\alpha^+ = r_{BA}^0 + d r_{BA}^2, \ \alpha^- = r_{AB}^0 + d r_{AB}^2 , \quad \beta^+ = r_{BA}^1 - r_{BA}^2, \ \beta^- = r_{AB}^1 - r_{AB}^2 .
\label{eq:alpha_beta_pm}
\end{equation}
Considering again the example of SIS epidemic, when $r_{AB}^0=\gamma$ and $r_{BA}^1 =\tau$  and the other rates are equal to zero, we have $\alpha^+ =0$, $\alpha^- =\gamma$, $\beta^+ =\tau$ and $\beta^- =0$. Hence the differential equation takes the form
\begin{equation}
\dot{a}_k = - \gamma a_k +\tau (1-a_k) \sum_{l=1}^N g_{kl} a_l .
\label{eq:SIS_NIMFA}
\end{equation}
Now we turn to the population-level approximation of the individual-level differential equation \eqref{eq:ak_regular}. This is a single differential equation instead of the system of $N$ equations and it is formulated in terms of the sum
$$
A(t) = \sum_{k=1}^N a_k(t) .
$$
The derivation of the population level approximation is based on the assumption that all variables $a_k$ in the individual-based system are equal to each other and then equal to $A/N$. We note that this is exact in the steady state when the graph is regular. Moreover, this is also exact if the initial conditions $a_k(0)$ are equal to each other and the graph is regular. Assuming $a_l=A/N$, the last sum in \eqref{eq:ak_regular} yields $\sum_{l=1}^N g_{kl} a_l = dA/N$ and then adding the equations in \eqref{eq:ak_regular} leads to
\begin{equation}
\dot{\overline{A}} = \alpha^{+} (N-\overline{A}) - \alpha^{-} \overline{A} + \left( \beta^{+} (N-\overline{A}) - \beta^{-} \overline{A} \right) \overline{A} \frac{d}{N},
\label{eq:A_popul}
\end{equation}
where we used $\overline{A}$ instead of $A$ to emphasise that this differential equation is based on an approximation and its solution $\overline{A}$ is not identical in general with $A$ which can be obtained by solving the system of $N$ differential equations \eqref{eq:ak_regular} and then summing the solutions $a_k$. Recall that $\overline{A}=A$ may hold, for example when the initial conditions $a_k(0)$ are equal to each other and the graph is regular.
	
The main result of the paper is that for graphs with special structure we are able to derive a numerically sharp estimate for the difference $\overline{A} - A$.

\section{Approximation error for regular Tur\'an graphs}
	
We will consider graphs with $N$ nodes partitioned into $M$ groups of equal size $m$, i.e. $N=Mm$, where all edges exist only between different groups and the vertices within the same group are not connected. If a node is connected to every other node outside its group then we have a complete multipartite graph with groups of equal size. In other words, this is a regular Tur\'an graph, which we denote by $T(N, m)$ or $T(Mm, m)$. While using both $M$ and $m$ is somewhat superfluous compared to writing $N/m$ instead of $M$, we prefer this more concise notation, and as $N$ denotes the size of all graphs we encounter this does not cause ambiguities. We note that the usual definition of Tur\'an graphs allows discrepancies of size 1 between group sizes, however, regularity is equivalent to the lack of such discrepancies. Tur\'an graphs were introduced and are primarily known due to their role in Tur\'an's theorem in extremal combinatorics \cite{Turan:1941}, with the simplest regular examples being the complete graph $T(N, 1)$ and the complete bi-partite graph $T(2m, m)$.

\subsection{Derivation of the error term} \label{sec:errorterm}
	
The advantage of considering regular Tur\'an graphs is that the sum $\sum\limits_{l=1}^N g_{kl} a_l$ in \eqref{eq:ak_regular} is the same for all indices $k$ that are in the same group. Now, we rewrite system \eqref{eq:ak_regular} for regular Tur\'an graphs by introducing the new variables $x_{ij} = a_k$ when $k$ is the $j$-th node in group $i$. Moreover, let
$$
A_i=\sum_{j=1}^m x_{ij}
$$
be the sum in the $i$-th group. Then
$$
\sum_{l=1}^N g_{kl} a_l =A-A_i,
$$
hence equation \eqref{eq:ak_regular} takes the form
\begin{equation}
\dot{x}_{ij} = \alpha^{+} (1-x_{ij}) - \alpha^{-} x_{ij} + \left( \beta^{+} (1-x_{ij}) - \beta^{-} x_{ij} \right) (A-A_i),
\label{eq:xij}
\end{equation}
for $j=1,2, \ldots, m$ and for $i=1,2, \ldots , M$, where note that $A$ can also be expressed as $A=A_1+A_2+\ldots +A_M$.
	
Now we can exploit the graph structure and use the linearity of the equations in $x_{ij}$ by adding them for $j=1,2, \ldots, m$ leading to
\begin{equation}
\dot{A}_{i} = \alpha^{+} (m-A_i) - \alpha^{-} A_i + \left( \beta^{+} (m-A_i) - \beta^{-} A_i \right) (A-A_i),
\label{eq:Ai}
\end{equation}
for $i=1,2, \ldots , M$. That is we have a self-contained system for the group variables $A_i$. In order to derive the error term of the population-level equation it is useful to introduce the parameters
\begin{equation}
\alpha = \alpha^{+} + \alpha^{-}, \quad \beta = \beta^{+} + \beta^{-}
\label{eq:alpha_beta}
\end{equation}
and rewrite \eqref{eq:Ai} as
\begin{equation}
\dot{A}_{i} = \alpha^{+} m - \alpha A_i + \left( \beta^{+} m - \beta A_i \right) (A-A_i),
\label{eq:Ai_new}
\end{equation}
Adding these differential equations for $i=1,2, \ldots , M$ and using $N=Mm$ yields
$$
\dot A = \alpha^{+} N - \alpha A + \beta^{+} (N-m) A  - \beta A^2 +\beta \sum_{i=1}^M A_i^2
$$
that can be rearranged by adding and subtracting the term $\beta A^2/M$ to
$$
\dot A = \alpha^{+} N - \alpha A + \beta^{+} (N-m) A  - \beta A^2 + \beta \frac{A^2}{M} +\beta \left( \sum_{i=1}^M A_i^2 - \frac{A^2}{M}\right) .
$$
Using $d=(M-1)m$ this equation can be written to a similar form as \eqref{eq:A_popul} for $\overline{A}$ as
\begin{equation}
\dot A = \alpha^{+} N - \alpha A + dA \left( \beta^{+}  - \beta \frac{A}{N}\right) +\beta H,
\label{eq:AH}
\end{equation}
where
\begin{equation}
H= \sum_{i=1}^M A_i^2 - \frac{A^2}{M} = \sum_{i=1}^M A_i^2 - \frac{1}{M} \left( \sum_{i=1}^M A_i \right) ^2
\label{eq:H_def}
\end{equation}
will be considered as the error term between the differential equation of $A$ and $\overline{A}$.
	
In the next subsection we derive an estimate on $H$ and then based on this we derive an estimate for the difference $\overline{A} - A$.

\subsection{Estimation of the error term H} \label{sec:error_estim}
	
The estimation of $H$ is based on the following lemma.
	
\begin{lemma}
Let $K\geq 2$ be an integer and let $c_k\in\mathbb{R}$ be arbitrary numbers for $k=1,2, \ldots, K$. Then the following identity holds
$$
K\sum_{k=1}^K c_k^2-\left(\sum_{k=1}^K c_k\right)^2=\sum_{k=1}^{K-1}\sum_{l=k+1}^{K}\left(c_k-c_l\right)^2=:\sum_{k<l}\left(c_k-c_l\right)^2.
$$
\end{lemma}
The statement can be proved by simply comparing the coefficients of the corresponding terms on the two sides of the equation, or by observing that both of them equals $K^2$ times the variance of a uniformly distributed random variable over the points $c_1, \dots, c_n$, according to the definition and a folklore identity in probability theory.
	
By using this lemma, we get
$$
H = \sum_{i=1}^{M} A_i^2 - \frac{A^2}{M}=\frac{1}{M}\sum_{i<j}\left(A_i-A_{j}\right)^2 \geq 0.
$$
This form of the error term $H$ shows that it is equal to zero if the number of $A$ nodes is the same in each group, that is $A_i=A_j$ for any two groups $i$ and $j$. Otherwise we have $H>0$, that is we have a lower bound on $H$.  Thus, \( H \) measures the deviation from a uniform distribution of $A$ nodes across the groups: the larger is \( H \), the more heterogeneous is the level of $A$ nodes among the groups.
	
Now we derive a differential inequality for $H$ to get an upper bound as well.
Differentiating the error term \( H \) yields
\[
\dot{H} = \frac{2}{M}\sum_{i < j} (A_i - A_j)(\dot{A}_i - \dot{A}_j).
\]
Using \eqref{eq:Ai_new} we get
\begin{align*}
\dot{A}_i - \dot{A}_j &= -(\alpha + \beta^+ m) (A_i-A_j) - \beta A (A_i-A_j)+ \beta \left(A_i^2 - A_j^2\right) \\ &= -(\alpha + \beta^+ m) (A_i-A_j) - \beta A_{ij} (A_i-A_j),
\end{align*}
where $A_{ij} = A- A_i-A_j\geq 0$. Hence
\[
\dot{H} = -\frac{2}{M}  \sum_{i < j} (A_i - A_j)^2(\alpha + \beta^+ m + \beta A_{ij})
\]
leading to the differential inequality
\[
\dot{H} \leq -2(\alpha + \beta^+ m)H .
\]
Integrating this inequality yields the following upper bound for $H$.
	
\begin{lemma} \label{Lemma_H_error_ineq}
Let $G=T(Mm, m)$ be a $d$-regular Tur\'an graph and let the parameters of the simple contagion be defined by \eqref{eq:alpha_beta_pm} and \eqref{eq:alpha_beta}. Then the error term $H$ defined in \eqref{eq:H_def} satisfies the inequalities
\begin{equation}
0\leq H(t) \leq H(0) e^{-bt}
\label{eq:H_estimate}
\end{equation}
for $t\geq 0$, where $b=2(\alpha + \beta^+ m)$.
\end{lemma}
The inequality shows that the differences between the number of $A$ nodes in different groups diminish exponentially over time.
	
We will need an upper bound on $H(0)$ as well, which is derived below.  Since all $A_i$ are non-negative, we have that
$$
\sum_{i=1}^{M} A_i^2 \leq \left( \sum_{i=1}^{M} A_i \right)^2
$$
Hence the definition \eqref{eq:H_def} of $H$ yields
$$
H=\sum_{i=1}^M A_i^2 - \frac{1}{M} \left( \sum_{i=1}^M A_i \right) ^2  \leq \frac{M-1}{M} \left( \sum_{i=1}^M A_i \right) ^2 = \frac{d}{N} \left( \sum_{i=1}^M A_i \right) ^2,
$$
where we used that for our graph we have $N=Mm$ and $d=(M-1)m$. Applying this inequality at time $t=0$ and using that $A=\sum A_i$, we obtain
\begin{equation}
N H(0) \leq dA^2(0) .
\label{eq:H0_estimate}
\end{equation}

\section{Estimate for the difference of the individual and population-level models} \label{sec:theor_estim}
	
The derivation of the upper bound is based on the fact that the differential equation \eqref{eq:AH} of $A$ can be considered as a perturbation of \eqref{eq:A_popul}, the differential equation of $\overline{A}$.
	
\subsection{A perturbation result for differential equations of Bernoulli type}
	
First, we prove the following lemma that can be used to derive the desired upper bound on $|A(t)-\overline{A}(t)|$.
	
\begin{lemma} \label{lem2}
Let $r_1\geq 0$ and $r_2, p, b, c>0$ and let $y$ be a positive solution of the differential equation
$$
\dot y (t) = p(r_1 +y(t)) (r_2-y(t)) + c \mbox{e}^{-bt} .
$$
Let $z$ be a positive solution of the unperturbed differential equation with $c=0$, that is
$$
\dot z(t) = p(r_1+ z(t)) (r_2-z(t))
$$
holds. Assume that the initial condition satisfies the inequality $c \leq p(r_1+z(0))^2$ and $y(0)=z(0)>0$. Then the following estimate holds for the relative difference of the perturbed and unperturbed solution for all $t>0$
$$
0\leq \frac{y(t) -z(t)}{z(t)} \leq  \frac{c (r_1+z(t))K(t)}{1-c(r_1+z(t))K(t)} \left( 1+ \frac{r_1}{z(t)} \right),
$$
where
\begin{equation}
K(t)=\exp(-at) \displaystyle \int_0^t \frac{\exp((a-b) s)}{(r_1+z(s))^2} ds , 
\label{eq:K}
\end{equation}
$$
r_1+z(t) = \frac{(r_1+r_2)(r_1+z(0))}{(r_2-z(0))\mbox{e}^{-at} +r_1+z(0)}
$$
and $a=p(r_1+r_2)$.
\end{lemma}

Note that even if $r_1=0$, $r_1+z(s)\geq \min(z(0), r_2)$, hence the integral defining $K(t)$ is finite for any $t$.

\begin{proof}[Lemma \ref{lem2}]
		
The differential equation of $z$ is of Bernoulli type, hence it can be solved by introducing the new unknown function $u=\frac{r_2-z}{r_1+z}$ that satisfies the linear differential equation
$$
\dot u(t) = -\dot z(t) \frac{r_1+r_2 }{(r_1+z(t))^2} = - p(r_2-z(t)) \frac{r_1+r_2 }{r_1+z(t)}  =-a u(t).
$$
Its solution is $u(t)= u(0) \mbox{e}^{-at}$ leading to
$$
z(t)=\frac{r_2-r_1 u(0) \mbox{e}^{-at}}{u(0) \mbox{e}^{-at}+1}  \textrm {, where } \quad u(0)=\frac{r_2-z(0)}{r_1+z(0)}.
$$
In the case of the function $y$ we introduce $v=\frac{r_2-y}{r_1+y}$ in a similar way. Its derivative is
\begin{align*}
\dot v(t) &= -\dot y(t) \frac{r_1+r_2 }{(r_1+y(t))^2} = - p(r_2-y(t)) \frac{r_1+r_2 }{r_1+y(t)} -  c \mbox{e}^{-bt} \frac{r_1+r_2 }{(r_1+y(t))^2} \\ &= -a v(t) -  c \mbox{e}^{-bt} \frac{r_1+r_2 }{(r_1+y(t))^2}.
\end{align*}
Let us introduce the difference of the solutions of the perturbed and unperturbed equations as $w=u-v$ and derive its differential equation.
\begin{align*}
\dot w(t) = \dot u(t) - \dot v(t) &= -a (u(t)-v(t)) +c \mbox{e}^{-bt} \frac{r_1+r_2 }{(r_1+y(t))^2} \\ &= -a w(t) +c \mbox{e}^{-bt} \frac{r_1+r_2 }{(r_1+y(t))^2}.
\end{align*}
We can consider this as an inhomogeneous linear differential equation. Hence multiplying by $\exp(at)$, then integrating and exploiting the initial condition $w(0)=0$ yields
$$
w(t) \exp(at) = c(r_1+r_2)   \int_0^t \frac{\exp((a-b) s)}{(r_1+y(s))^2} ds .
$$
Using the identity
$$
w=u-v= \frac{r_2-z}{r_1+z} - \frac{r_2-y}{r_1+y}=\frac{(r_1+r_2)(y-z)}{(r_1+z)(r_1+y)}
$$
leads to
$$
\frac{r_1+r_2}{(r_1+z(t))(r_1+y(t))}(y(t)-z(t))\exp(at) = c(r_1+r_2)  \int_0^t \frac{\exp((a-b) s)}{(r_1+y(s))^2} ds
$$
therefore
$$
y(t)-z(t) = c (r_1+z(t))(r_1+y(t))\overline{K}(t),
$$
where we introduced
$$
\overline{K}(t):=\exp(-at) \displaystyle \int_0^t \frac{\exp((a-b) s)}{(r_1+y(s))^2} ds.
$$
An upper bound will be derived for the function $\overline{K}(t)$. The perturbation is positive, the initial conditions are identical hence the differential equations of $z$ and $y$ imply that $z(t)\leq y(t)$ for all $t>0$. This yields $\overline{K}(t)\leq K(t)$, where $K(t)$ is defined in \eqref{eq:K}.
Using that $c>0$ we obtain
$$
r_1+y(t)-(r_1+z(t)) = y(t)-z(t)\leq c (r_1+z(t))(r_1+y(t))K(t).
$$
Expressing $r_1+y(t)$ yields
\begin{equation}
r_1+y(t)\leq \frac{r_1+z(t)}{1-c(r_1+z(t))K(t)},
\label{eq:ineq_y_z}
\end{equation}
assuming that $1-c(r_1+z(t)) K(t)>0$. Now we prove that the assumption $c \leq p(r_1+z(0))^2$ ensures that this inequality holds.
		
Using the formulas for the functions $z$ and $K$ we need to prove that
$$
\frac{c (r_1+r_2)}{u(0)  \exp(- a t)+1} \cdot \exp(-at) \displaystyle \int_0^t \exp((a-b) s) \frac{(u(0) \exp(- a s)+1)^2}{(r_1+r_2)^2} ds <1 .
$$
After simple algebra, multiplying by $\exp(at)(u(0)  \exp(- a t)+1)$ one obtains that the desired inequality takes the form
$$
\frac{c}{r_1+r_2}  \displaystyle \int_0^t \exp((a-b) s)(u(0) \exp(- a s)+1)^2 ds < (u(0)  \exp(- a t)+1)\exp(at) .
$$
Consider the functions in the left and right hand sides. Their values at $t=0$ satisfy the inequality $0 < u(0)+1$ since $u(0)+1=\frac{r_1+r_2}{r_1+z(0)}>0$. Hence it is enough to prove that their derivatives satisfy the corresponding inequality
$$
\frac{c}{r_1+r_2}   \exp((a-b) t)(u(0) \exp(- a t)+1)^2 \leq a\exp(at).
$$
The left hand side can be estimated from above with $\frac{c}{r_1+r_2}   \exp(a t)(u(0)+1)^2 $, thus we need the estimate
$$
\frac{c}{r_1+r_2}  (u(0)+1)^2  \leq a .
$$
Using that $u(0)+1=\frac{r_1+r_2}{r_1+z(0)}$ and $a=p(r_1+r_2)$ we get
$$
\frac{c}{r_1+r_2}  \frac{(r_1+r_2)^2}{(r_1+z(0))^2}  \leq p(r_1+r_2)
$$
that is ensured by the assumption $c \leq p(r_1+z(0))^2$. Thus we have proved that  $1-c(r_1+z(t)) K(t)>0$.
		
Since we have that $z(t)\leq y(t)$ for all $t>0$, inequality \eqref{eq:ineq_y_z} yields the statement that we wanted to prove
\begin{align*}
y(t) -z(t) &= r_1+ y(t) -(r_1+z(t)) \\ & \leq \frac{r_1+z(t)}{1-c(r_1+z(t))K(t)} -(r_1+z(t)) = (r_1+z(t)) \frac{c(r_1+z(t))K(t)}{1-c(r_1+z(t))K(t)}
\end{align*}
leading to
\begin{align*}
0 \leq \frac{y(t) -z(t)}{z(t)} \leq  \frac{c (r_1+z(t)) K(t)}{1-c(r_1+z(t))K(t)}  \left( 1+ \frac{r_1}{z(t)} \right).
\end{align*}
\end{proof}

\subsection{Upper bound on $|A(t)-\overline{A}(t)|$}

Applying the lemma above, we can estimate the difference $|A(t)-\overline{A}(t)|$ based on the upper bound \eqref{eq:H_estimate} ensuring that the error term is small enough. In order to apply the lemma, recall that $\overline{A}$ satisfies
$$\dot {\overline{A}} = \alpha^{+} N - \alpha \overline{A} + d\overline{A} \left( \beta^{+}  - \beta \frac{\overline{A}}{N}\right) .
$$
The quadratic polynomial in the right hand side has a negative and a positive root that will be denoted by $-r_1$ and $r_2$ respectively. Hence introducing $p=d\beta/N$, the differential equation above can be written as
$$
\dot {\overline{A}} = p(r_1+\overline{A})(r_2-\overline{A}) .
$$
Thus the differential equation \eqref{eq:AH} of $A$ reads as
$$
\dot {A} = p(r_1+A)(r_2-A)+\beta H .
$$
Now we are in the position of stating and proving the main result of the paper.
	
\begin{theorem}\label{theo:main}
Let $G=T(Mm, m)$ be a $d$-regular Tur\'an graph and let the parameters of the simple contagion be defined by \eqref{eq:alpha_beta_pm} and \eqref{eq:alpha_beta}. Let $r_1, r_2$ and $p$ be as defined above.
Let $a_k$ be a solution of system \eqref{eq:ak_regular}, $A=a_1+\ldots + a_N$ and $\overline{A}$ be the solution of \eqref{eq:A_popul} with initial condition $\overline{A}(0)= A(0)$. Then the following estimate holds for the relative difference for all $t>0$
$$
0\leq \frac{A(t)-\overline{A}(t)}{\overline{A}(t)} \leq   \frac{ \beta H(0) (r_1+\overline{A}(t)) K(t)}{1-\beta H(0) (r_1+\overline{A}(t))K(t)} \left(1+ \frac{r_1}{\overline{A}(t)}\right),
$$
where
$$
K(t)=\exp(-at) \displaystyle \int_0^t \frac{\exp((a-b) s)}{(r_1+\overline{A}(s))^2} ds,
$$
$$
r_1+\overline{A}(t) = \frac{(r_1+r_2)(r_1+A(0))}{(r_2-A(0))\mbox{e}^{-at} +r_1+A(0)}
$$
with $a=p(r_1+r_2)$ and $b=2(\alpha + \beta^+ m)$.
		
\end{theorem}
	
\begin{proof}
The functions $A$ and $\overline{A}$ satisfy the differential equations above and we have the bounds \eqref{eq:H_estimate} on $H$.
Let us introduce $A^*$ as the solution with the upper bound in the right hand side.
\begin{equation} \label{felsobecsl_de}
\dot {A^*} = p(r_1+A^*)(r_2-A^*)+\beta H(0)\exp( -b t) .
\end{equation}
The right hand side of the equation for $A$ is between those for $\overline{A}$ and $A^*$. Hence starting the solutions of the three differential equations from the same initial condition $\overline{A}(0)=A(0)=A^*(0)$, we have $\overline{A}(t)\leq A(t) \leq A^*(t)$ for all $t>0$.
		
We will estimate the difference of $\overline{A}$ and $A^*$ and this will yield the estimate of the difference of $\overline{A}$ and $A$.
Let us apply Lemma $\ref{lem2}$ with $z=\overline{A}$, $y=A^*$ and $c=\beta H(0)$. We get the desired statement if we check that $c \leq p(r_1 +\overline{A}(0))^2$. Using the above definitions of the constants, this inequality takes the form
$$
\beta H(0) \leq \beta\frac{d}{N} (r_1 +\overline{A}(0))^2 .
$$
Since $r_1\geq 0$ and $\overline{A}(0)=A(0)$, this inequality holds if $N H(0) \leq d A^2(0)$, which is true according to  \eqref{eq:H0_estimate}. This completes the proof.
\end{proof}

Note that the right hand side of the inequality in the theorem is an explicitly integrable function, that is the upper bound for the difference can be explicitly calculated once the parameters of the model are given. In the next section we will show examples of graphs for which this function is determined in the case of SIS epidemic and numerical evidence is given that this upper bound is rather sharp.
	
\section{Numerical investigation of the sharpness of the estimate in the case of epidemic spread} \label{sec:sharpness}

Now let us apply the result of Theorem \ref{theo:main} for the SIS model.
In the case of SIS epidemic propagation, when each node is either infected A = I or susceptible B = S, equation \eqref{eq:A_popul} is the well-known mean-field model \eqref{MF_NIMFA} and
 equation \eqref{eq:AH} takes the form
\begin{equation} \label{NIMFA_SIS}
\dot {I} = - \gamma I + d\tau I \left( 1  - \frac{I}{N}\right) +\tau H ,
\end{equation}

where $H= \sum_{i=1}^M I_i^2 - \frac{I^2}{M}$, since $d=m(M-1)$, $\alpha^{+}=0$, $\alpha=\alpha^{-}=\gamma$, $\beta^{-}=0$, $\beta=\beta^{+}=\tau$ (see \eqref{eq:alpha_beta_pm}, \eqref{eq:alpha_beta}).

According to Lemma \ref{Lemma_H_error_ineq} we have $0\leq H(t) \leq H(0) e^{-bt}$, where $$b=2(\tau m+\gamma) \textrm{ and } \ a=\tau d-\gamma,$$ since the remaining paramaters in Lemma \ref{lem2} are $r_1=0$, $r_2=N\big(1-\frac{\gamma}{\tau d} \big)$ and $p=\frac{\tau d}{N}$.

Although the estimate of Theorem \ref{theo:main} works with any initial condition, for simplicity, now we consider the initial condition when the infection starts from a single node, hence $I(0)=1=\overline{I}(0)$ yielding $H(0)=\frac{M-1}{M}$. The solution of the mean-field model \eqref{MF_NIMFA} can be easily obtained as
	$$
	\overline{I}(t)=\frac{N a}{(Na-\tau d)\exp(-at) +\tau d } \ , \quad \mbox{ with } d=m(M-1), \  a=\tau d-\gamma.
	$$
	Now, the integral in the formula of $K(t)$ can be explicitly calculated and takes the following form
	$$
	K(t) = \frac{\exp(-at)}{N^2a^2} \left( C_1(\mbox{e}^{-q_1 t} -1) + C_2(\mbox{e}^{-q_2 t} -1) + C_3(\mbox{e}^{-q_3 t} -1) \right),
	$$
	where
	$$
	q_1=\tau (N+m) +\gamma,\quad  q_2 = 3\gamma - \tau (N-3m), \quad  q_3 = 2(\tau m+\gamma) ,
	$$
	$$
	C_1= \frac{(N\gamma -(N-1)(N-m) \tau)^2}{-(N+m)\tau -\gamma}, \quad  C_2= \frac{(N-m)^2 \tau^2}{(N-3m)\tau -3\gamma}, $$
	$$\ C_3= \frac{\tau(N-m)(N\gamma -(N-1)(N-m) \tau)}{\tau m+\gamma} \ .
	$$
	Thus we have the following corollary of Theorem \ref{theo:main} for a Turán graph.
		
	\begin{corollary} \label{cor1}
		Consider the case of SIS epidemic propagation on a Turán graph $T(N,m)$ with $N=Mm$, $d=m(M-1)$, where the error term $H$ satisfies $0\leq H(t) \leq H(0) e^{-bt}$ with $b=2(\tau m+ \gamma)$. Let $a_k$ be a solution of \eqref{eq:SIS_NIMFA}, $I=a_1+\ldots + a_N$, i.e. $I$ satisfies equation \eqref{NIMFA_SIS}, and let $\overline{I}$ be the solution of the population-level model \eqref{MF_NIMFA}. Assume that the infection starts from a single node, i.e. $I(0)=1= \overline{I}(0)$. Then the following estimate holds for all $t>0$ for the relative difference
		$$
		0\leq \frac{I(t)-\overline{I}(t)}{\overline{I}(t)} \leq   \frac{ L(t)}{1-L(t)},
		$$
		where
		$$
		L(t)=\frac{\tau (M-1) \exp(-at)}{Na M((Na-\tau d)\exp(-at) +\tau d )} \sum_{i=1}^3 C_i(\mbox{e}^{-q_i t} -1),
		$$
        where the parameters are as defined above.
	\end{corollary}
	
\begin{figure}[htbp]
		\centering
		\setlength{\unitlength}{\textwidth}
		\begin{picture}(1,0.5)
		\put(-0.05,0){\includegraphics[width=1\unitlength]{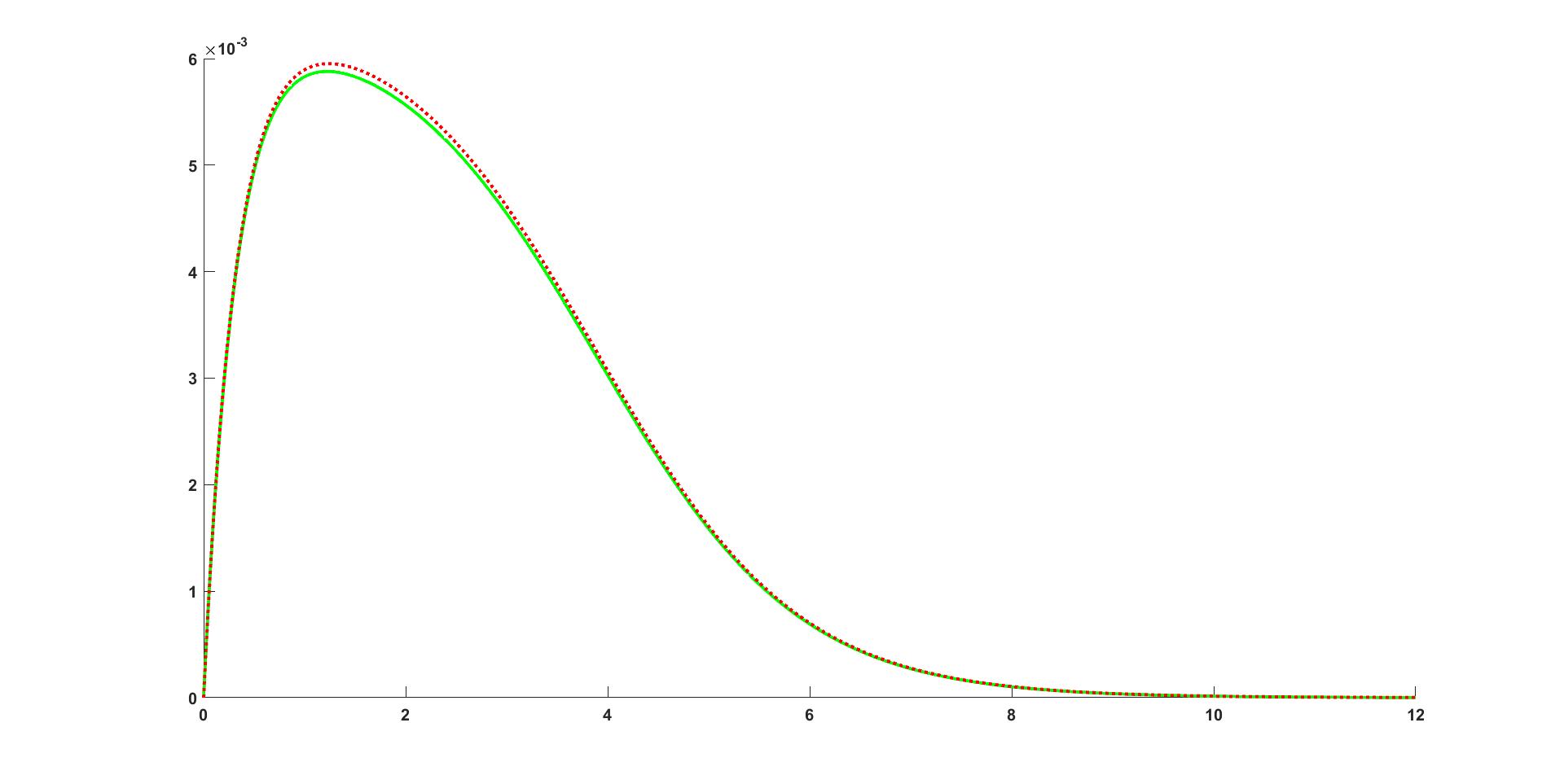}}
		\end{picture}
		\caption{The sharpness of the estimate of Corollary \ref{cor1} in the case of SIS epidemic propagation. The relative difference $\frac{I(t)-\overline{I}(t)}{\overline{I}(t)}$  (green) on a Turán graph $T(100,5)$  starting with one infected node and the upper bound (red) given by Corollary \ref{cor1}. The function $I$ satisfies equation \eqref{NIMFA_SIS} and $\overline{I}$ is the solution of the population-level model \eqref{MF_NIMFA}. The epidemic parameters are $\gamma=1$, $\tau=\frac{2\gamma}{d}=\frac{2}{95}$ (chosen in such a way that half of the nodes in the graph become infected asymptotically.)}
		\label{fig:100a}
	\end{figure}

In order to show the sharpness of the estimate given in the corollary, we plot the relative difference $\frac{I(t)-\overline{I}(t)}{\overline{I}(t)}$ and the upper bound given by Corollary \ref{cor1} for two types of Turán graphs. In Figure \ref{fig:100a}, the two curves corresponding to the Tur\'an graph $T(100,5)$, consisting of 20 groups of size 5, are shown. One can see that there is only a slight difference between the approximation error and its theoretical upper bound.

\begin{figure}[htbp]
		\centering
		\setlength{\unitlength}{\textwidth}
		\begin{picture}(1,0.5)
		\put(-0.05,0){\includegraphics[width=1\unitlength]{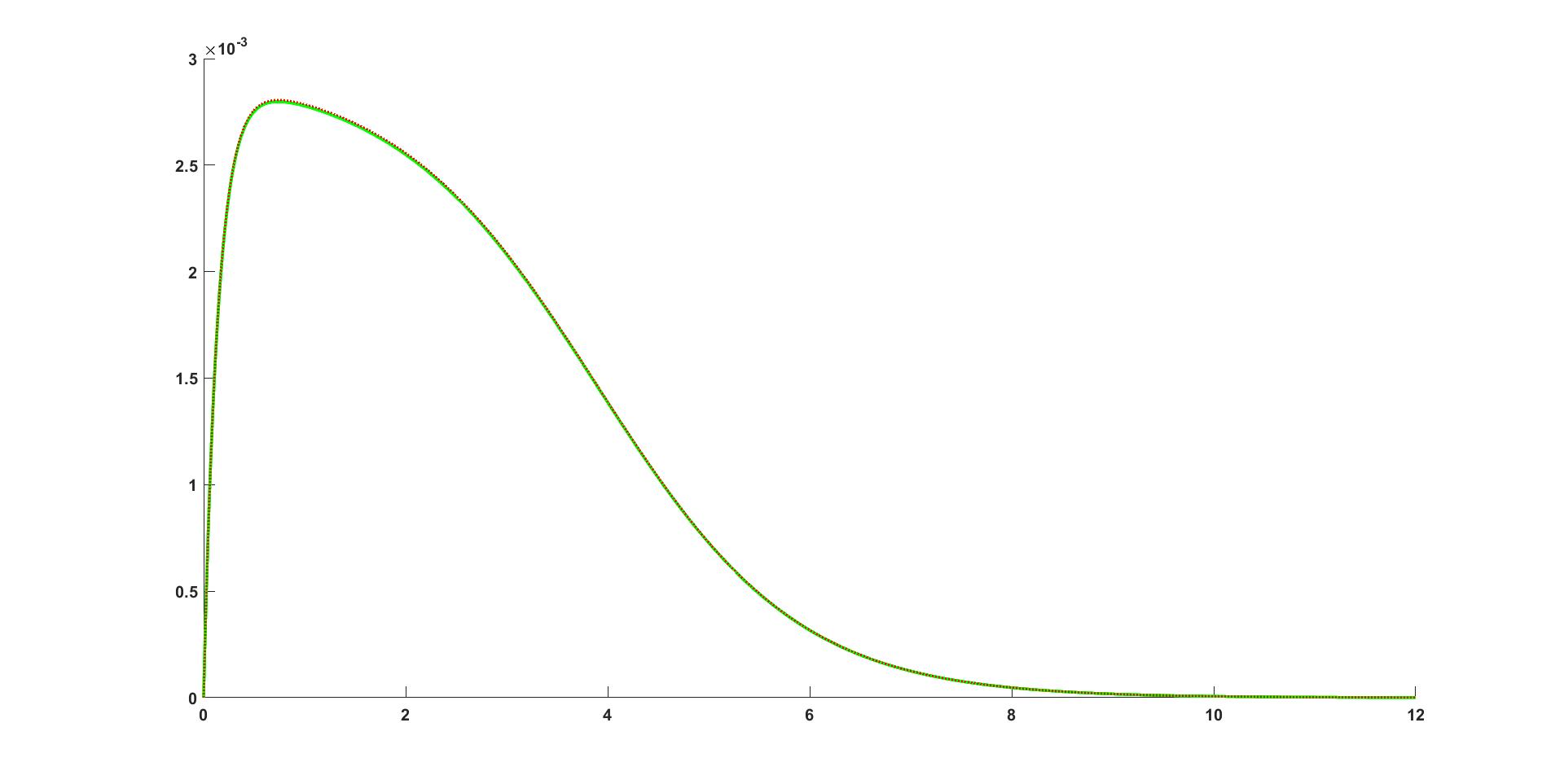}}
		\end{picture}
		\caption{The sharpness of the estimate of Corollary \ref{cor1} in the case of SIS epidemic propagation. The relative difference $\frac{I(t)-\overline{I}(t)}{\overline{I}(t)}$  (green) on a complete bi-partite graph with $N=100$ nodes starting with one infected node and the upper bound (red) given by Corollary \ref{cor1}. The function $I$ satisfies equation \eqref{NIMFA_SIS} and $\overline{I}$ is the solution of the population-level model \eqref{MF_NIMFA}. The epidemic parameters are $\gamma=1$, $\tau=\frac{2\gamma}{d}=\frac{2}{50}$ (chosen in such a way that half of the nodes in the graph become infected asymptotically.)}
		\label{fig:100b}
	\end{figure}

    Figure \ref{fig:100b} shows the case of the complete bi-partite graph, $T(100,50)$.  Here the upper bound is extremely sharp, the difference between the approximation error and its upper bound is negligible.

	The significance of our result can be briefly summarized as follows. It is enough to solve a single differential equation for $\overline{I}(t)$ instead of solving a large system of differential equations for $a(t)$, and this yields us the prevalence $I(t)$ with high accuracy. Moreover, an explicit upper bound on the accuracy is given by Corollary \ref{cor1}.

\section{Discussion}

In this paper, we studied the individual-based model of simple contagions on regular graphs, exploiting the fact that this system of differential equations explicitly includes the adjacency matrix of the network. A one-dimensional approximation, called population-level model, was introduced and compared to the full model. For many networks, this single differential equation yields an excellent approximation of the total amount of nodes in a given status, despite of the fact, that exact value of this quantity is determined by a system of $N$ differential equations, where $N$, the number of nodes in the network, can be very large. A theoretical estimate is derived for the difference of the individual and population-level model for regular Tur\'an graphs, which proves to be rather sharp for dense graphs. According to our knowledge, this is the first theoretical result on the accuracy of the population-level approximation.

The investigations of this paper can be extended to several directions. On one hand, the theoretical estimate for the difference of the solutions of the one-dimensional equation and the full system could be proved not only for Tur\'an graphs but for dense regular graphs in general. Our method of proof unfortunately does not extend to this general class directly, because a crucial step of the proof, the exponential estimate for the error term $H$ does not hold even for slight modification of a Tur\'an graph as it can be easily checked even for a bi-partite graph with six nodes. On the other hand, a two-dimensional approximation could be derived, containing not only the total quantity $A$ but also the correction term $H$.

All the investigations in this paper correspond to regular graphs, hence an obvious demand arises for having similar results in the cases of graphs with a more complex degree distribution. As a first step in this direction one can consider bimodal graphs that have two different degrees. The nodes are divided into two groups, in the first one there are $N_1$ nodes with degree $d_1$ and in the second group there are $N_2$ nodes of degree $d_2$. We note that this degree distribution can be realized with several different adjacency matrices, thus further properties of the graphs will certainly have effect on the performance of the approximation. The one-dimensional approximation is again \eqref{MF_NIMFA}, with $N=N_1+N_2$ and average degree $d=(N_1d_1+N_2d_2)/N$. However, it will perform poorly, hence a system of two variables, tracking the number of $A$ nodes in the two degree classes separately, is a candidate as a population-level approximation. The theoretical estimate for the accuracy of this two-dimensional approximation could be another extension of the result in this paper.


\section*{Acknowledgements}

The study was funded by the National Research, Development and Innovation Office in Hungary (RRF-2.3.1-21-2022-00006).
	
P.L. Simon acknowledges support from the Hungarian Scientific Research Fund, OTKA (grant no. 135241).

B. Maga acknowledges support from the KKP 139502 project, funded by the Ministry of Innovation and Technology of Hungary from the National Research, Development and Innovation Fund.



\end{document}